\documentclass[Afour,sageh,times]{sagej}

\usepackage{moreverb,url}

\usepackage[colorlinks,bookmarksopen,bookmarksnumbered,citecolor=red,urlcolor=red]{hyperref}

\newcommand\BibTeX{{\rmfamily B\kern-.05em \textsc{i\kern-.025em b}\kern-.08em
		T\kern-.1667em\lower.7ex\hbox{E}\kern-.125emX}}

\def\volumeyear{2016}

\begin{document}
	
	\runninghead{Beall et al.}
	
	\title{Design Considerations for Factorial Adaptive Multi-Arm Multi-Stage (FAST) Clinical Trials }
	
\author{
		Jonathan Beall\affilnum{1},
		Jordan Elm \affilnum{1}, 
		Mathew W Semler \affilnum{2}, 
		Li Wang \affilnum{3},
		Todd Rice\affilnum{2}, 
		Hooman Kamel \affilnum{4}, 
		William Mack \affilnum{5}, 
		Akshitkumar M. Mistry \affilnum{6}
	}

	\affiliation{
		\affilnum{1} Department of Public Health Sciences, Medical University of South Carolina, Charleston, South Carolina\\
		\affilnum{2} Department of Medicine, Division of Allergy, Pulmonary, and Critical Care Medicine, Vanderbilt University Medical Center, Nashville, Tennessee\\
		\affilnum{3} Department of Biostatistics, Vanderbilt University School of Medicine, Nashville, Tennessee\\
		\affilnum{4} Department of Neurology, Weill Cornell Medicine, New York, New York\\
		\affilnum{5} Department of Neurosurgery, Keck School of Medicine, University of Southern California, Los Angeles, California \\
		\affilnum{6} Department of Neurosurgery, University of Louisville, Louisville, Kentucky \\
	}
	
	\corrauth{Jonathan Beall
		Department of Public Health Sciences, 
		Medical University of South Carolina,
		Charleston, South Carolina, U.S.A}
	
	\email{bealljo@musc.edu}
	
	\begin{abstract}

	Multi-Arm, Multi-Stage (MAMS) clinical trial designs allow for multiple therapies to be compared across a spectrum of clinical trial phases. MAMS designs can be categorized into several overarching design groups, including adaptive designs (AD) and multi-arm (MA) designs. Factorial clinical trials designs represent an additional group of designs which can provide increased efficiency relative to fixed, traditional designs. In this work, we explore design choices associated with Factorial Adaptive Multi-Arm Multi-Stage (FAST) designs, which represent the combination of factorial and MAMS designs. This category of trial can potentially offer benefits similar to both MAMS and factorial designs. This work is motivated by a proposed clinical trial under development. 
	
	\end{abstract}
	
	
	\maketitle
	
	\section{Introduction}
	
	Multi-Arm, Multi-Stage (MAMS) are an increasingly popular class of clinical trials designs which provide a framework where several therapies can be assessed across multiple clinical trial phases \citep{Ghosh2020,Lin2017}.  Generally falling under the categories of adaptive designs (AD) and multi-arm (MA) designs, MAMS often provide increases in efficiency relative to traditional trial designs by allowing for multiple active treatment arms to be compared against a single shared control  \citep{Lin2017,Wason2012}.

	Like MAMS designs, factorial clinical trials are another class of trial design which offer increases in efficiencies relative to traditional design by allowing for multiple treatment arms to be compared against a single control arm \citep{White2022,Jaki2016}.  In order to achieve these efficiencies, factorial design often assume that there is no interaction effect present among the active treatment arm, with this claim of no interaction often needing some form of prior support \citep{Sydes2009,Jaki2016,White2022}.   
	
	Factorial Adaptive Multi-Arm Multi-Stage (FAST) designs represent the combination of factorial and MAMS design and offer benefits similar both of its’ predecessors, namely the ability to compare multiple active treatment arms to a single shared control under a framework which spans multiple trial phases and allows for pre-specified adaptations \citep{White2022}. While MAMS and factorial designs have each been thoroughly researched and utilized, a review of the literature suggest that FAST designs are both understudied and underutilized.  Previous work by White et. al (2022) explored a general framework for FAST (or factorial-MAMS) designs; however, their work focused on the practical advantages/disadvantages of the FAST design. 
	
	The purpose of this work is to build on the current literature by exploring the impact of the timing and type of assessments during the Phase II portion of a seamless Phase II/III FAST design on the final operating characteristics. While this work is motivated by a clinical trial which is currently under development, detailed below, the results will provide insight into the many aspects which need to be carefully considered during the design of a FAST clinical trial.

	\section{Motivating Trial}
	\label{s:model}
	
	This work is motivated by a multicenter, factorial, randomized clinical trial currently being planned to compare the effectiveness of different fluid therapies and a mineralocorticoid in the treatment of patients with subarachnoid hemorrhage from spontaneous rupture of an intracranial aneurysm. Specifically, two newer fluid therapies will be compared to the standard use of saline (control), and the use of mineralocorticoid will be compared to no use (control). The primary objective is to determine whether any of the newer fluid therapies and the use of mineralocorticoid improves the outcomes of patients with subarachnoid hemorrhage. The primary outcome of the study is the proportion of patients who are assessed to have a functional outcome score of 4 to 6, which corresponds to poor outcomes on an ordinal scale (called modified Rankin Scale, or mRS) that ranges from 0 to 6. 
	
	
	Given the goal of testing two therapeutic domains, two newer fluid therapies and a mineralocorticoid, a 3x2 factorial trial was initially selected as an ideal design.  However, the challenging aspect of this trial stems from the fluid intervention rather than the mineralocorticoid. Because of the acute and often unexpected changes experienced by subarachnoid hemorrhage patients that often lead to treatment with more than one fluid, “contamination” with other fluids in practice raises concerns about trial feasibility for the fluid interventions. Therefore, a seamless phase 2/3 testing is implemented for the fluid factors, while the mineralocorticoid simultaneously undergoes only phase 3 testing. This combination leads to a FAST design being optimal for the proposed questions of interest. 
	
	For the three fluids being tested, each have distinct electrolyte composition which generate distinct physiological effects in patients. For example, saline has the lowest bicarbonate and highest sodium content among the 3 fluids; therefore, fluid therapy with saline leads to lower plasma bicarbonate levels when compared to therapy with the newer fluids. We hypothesize that the therapeutic effect of fluids is mediated by changes in bicarbonate or sodium levels in the plasma. As such, bicarbonate and sodium can be considered outcomes of interest in the Phase II portion of the trial.  Utilizing bicarbonate and sodium as biologic markers of therapeutic effect offers an opportunity for two assessments in the Phase II portion of the trial to increase overall trial efficiency. 
	
	An adaptation in Phase II would be to drop randomization into one of the newer fluids.  Dropping one of the two newer fluid therapies is very desirable to conserve resources in Phase III testing. Whichever of the two newer fluids that generates the greatest effect in the plasma bicarbonate and sodium in the opposite direction compared to the expected levels produced by treatment with saline will be advanced. For example, the prior expectation is that, among the 3 fluid therapies considered, treatment with saline will result in the lowest bicarbonate levels. As such, the goal of the arm dropping assessment would be to compare the two novel fluid therapies to assess for which therapy produces the greatest bicarbonate levels, ultimately creating the greatest effect of treatment when compared to saline on bicarbonate levels.  In the absence of a clear differences between the novel fluid therapies in both effects, one pre-selected newer therapy will be advanced by default. In the event that at least one significant difference is observed when comparing the novel fluid therapies,  decision rules are explicitly defined for retaining either one or both of the therapies.
	
	A second potential assessment in Phase II would be to determine the feasibility of the trial. Since the prior expectation is that the electrolyte composition of the included fluids mediates changes patient to outcomes, if the expected differences when comparing the novel fluid therapies to saline on the electrolyte compositions is not observed, then the overall fluid assessment can be terminated. 
	 
	Given the distinct goals for the fluid and mineralocorticoid domains, a FAST design is best suited to address all hypotheses to be tested and allow for increased efficiency relative to more traditional designs.

\section{Study Design}
Following our motivating example, we explore a seamless Phase II/III FAST design which includes 2 interim assessments and a final assessment. The primary aim of this work is to examine the impact of the timing for the interim assessments on final operating characteristics. The interim analyses will constitute the Phase II portion of the trial with the final assessment representing the Phase III portion of the trial. Sample sizes for the interim analysis and final operating characteristics will be determined via simulation.  

To accommodate the clinical aims of the hypothesized clinical trial, we propose a trial design which allows for multiple questions of interest to be jointly assessed through a FAST design. Henceforth, we will reference each question of interest as a domain.  In general, we let d=A,B,…$\Delta$ reference the domains of interest with $\Delta$ representing the maximum number of domains.  From our motivating trial, an example of a domain would be the comparison of the newer fluid arms against the saline control arm.  
Within in each domain, we will note treatment assignment within domain as $d_t$  where $t=1,2,…\tau$ and $\tau$ represents the maximum number of treatment arms in domain $d$.  Again utilizing our motivating trial as an example, letting the first domain, $d=A$, indicate the comparison of the newer fluid therapies to saline, there would then be three treatment arms in this domain, $A_1=Saline$, $A_2=Fluid_1$, and $A_3=Fluid_2$. 

For the outcomes of interest, we will note outcomes by trial phase. Thus, let $Y_{po}$ represent the $o^{th}$ outcome of interest for the $p^{th}$ trial phase, where $o=1,2,…,\theta$ and $p=1,2,…,\rho$. From our motivating example, the outcomes utilized in the Phase II portion of the trial represent biomarkers assumed to be affected by treatment and the Phase III outcome of interest represents a functional outcome. We will assume two analyses within the Phase II portion of the trial and one in the Phase III portion of the trial.  This is done to mirror the proposed design for the motivating trial.  The Phase II portion of the trial will consist of 2 analyses: a feasibility analysis and an arm-dropping analysis. 

\subsection{Phase II}

\subsubsection{Arm Dropping Analysis}
The goal of the arm-dropping analysis is to determine which, if any, of the treatment arms (novel fluid therapies) can be discontinued from enrollment. We let $\mu_{A_1}^{Y_{11}}$ and $\mu_{A_2}^{Y_{11}}$ represent the average value for outcome $Y_{11}$ (bicarbonate) for subjects in arms 1 and 2 in Domain A, respectively. Further, we let $\mu_{A_1}^{Y_{12}}$ and $\mu_{A_2}^{Y_{12}}$represent the average value for outcome $Y_{12}$ (sodium) for subjects in arms 1 and 2 in Domain A, respectively. Then, utilizing independent t-tests by outcome, the non-control arms will be compared by assessing two hypotheses:

$$H_0:\mu_{A_1}^{Y_{11}}=\mu_{A_2}^{Y_{11}} \hspace{2mm}  vs.\hspace{2mm}  H_a:\mu_{A_1}^{Y_{11}} \neq \mu_{A_2}^{Y_{11}}$$ 
$$H_0:\mu_{A_1}^{Y_{12}}=\mu_{A_2}^{Y_{12}} \hspace{2mm} vs.\hspace{2mm}  H_a:\mu_{A_1}^{Y_{12}} \neq \mu_{A_2}^{Y_{12}}$$

The results of these hypothesis tests will be used to determine which arms will continue to enroll. If there is a statistically significant difference observed for outcome $Y_{11}$, the treatment arm in Domain A with the highest average outcome $Y_{11}$  value will be retained. If there is a difference in outcome $Y_{12}$ between the treatment arms in Domain A, then the arm with the lowest outcome $Y_{12}$ average will be retained. There is a possibility that neither arm will be dropped. 

\subsubsection{Feasibility Analysis}
The goal of the feasibility analysis is to assess, utilizing one or more biomarkers, if the expected effect of treatment is being observed on the biomarker(s). While the comparison for this example focuses on responses in key biomarkers, any appropriate Phase II assessment could be included in this analysis. Treatment arms will be compared using an independent one-sided t-test. Letting $\mu_{A_0}^{Y_{11}}$ represent the average value for outcome $Y_{11}$ for subjects in arm $A_0$ (the control arm for Domain A) and $\mu_{A_{(1+2)}}^{Y_{11}}$ represent the average outcome $Y_{11}$ for the pooled treatment arms from Domain A, we then propose a one-sided test, with hypothesis below:

$$H_0:\mu_{A_0}^{Y_{11}} \geq \mu_{A_{(1+2)}}^{Y_{11}} vs.H_a:\mu_{A_0}^{Y_{11}} \leq \mu_{A_{(1+2)}}^{Y_{11}} $$

If statistical significance is not obtained, then the assessment of treatment effect within domain A will terminate and subjects will no longer randomized to any of the treatments within domain A. 
 
For this analysis, all subject data are utilized in the comparison. Pooling data within $A_1$ and $A_2$ could result in an assessment which is under-powered if there is a differential effect of treatment on $Y_{11}$. Given this pooling, in addition to the treatment effect, the impact of this pooling is likely affected by the timing and ordering of the arm-dropping and feasibility assessments.  The impact of these trial characteristics and their impact on overall operating characteristics will be assessed via simulation.  

\subsection{Phase III}
For any trial where domain A proceeds beyond the Phase II portion of the trial, the final analysis will consist of an assessment comparing each treatment arm to control within domain.  The final analysis will be conducted by constructing a generalized linear model where the outcome of interest is the Phase III outcome and covariates in the model represent the treatment assignments for each subject.  An appropriate correction for the multiple comparisons will be applied to $\alpha$ so that Type 1 Error is controlled \citep{Wang2009}. Due to the arm dropping analysis, there are three possible scenarios for the final analysis: where a treatment in domain A is dropped, where no treatment in domain A is dropped, where domain A is terminated. 

\subsubsection{Domain A Continued with One Arm Retained}
In this scenario, while the majority of subjects will be randomized to one of the newer fluid arms or saline, a small proportion of subjects will be randomized to the dropped fluid arm prior to the arm dropping analysis.  For the final analysis, subjects will be grouped as either having been randomized to saline or one of the newer fluid arms; that is, subjects randomized to either of the non-saline arms throughout the duration of the trial are pooled into a single fluid therapy arm for the final analysis. 

Therefore, letting $I(X_{A_0 i}=1)$ represent an indicator variable for if subject $i$ was randomized to saline, $I(X_{B_1 i}=1)$ represent an indicator variable for if subject $i$ was randomized to fludrocortisone, and $p_i = P(Y_{21}=1)$, we can then construct the primary analysis model as below. 

$$logit(p_i)=\beta_0+\beta_1 I(X_{A_0 i}\neq1)+\beta_2 I(X_{B_1 i}=1)$$

To control the family-wise type I error in the strong sense at $\alpha=0.05$, we will then assess for any significant treatment effect among the fluid and fludrocortisone arms using a gatekeeping procedure: 

$$H_0:\beta_1=\beta_2=0 \hspace{2mm} vs.\hspace{2mm}  H_a:\beta_1\neq 0|\beta_2\neq0$$

If the null hypothesis for the test above is rejected, implying that either the pooled fluid arm or the fludrocortisone arm has a non-zero treatment effect relative to the saline/no fludrocortisone arm, then individual tests comparing each arm to control will be completed. 

$$H_0:\beta_1=0 \hspace{2mm} vs. \hspace{2mm} H_a:\beta_1 \neq 0$$
$$H_0:\beta_2=0 \hspace{2mm} vs. \hspace{2mm} H_a:\beta_2 \neq 0$$

As with the feasibility assessment in Phase II, it is worth noting that the comparisons above utilize the pooling approach so that all patient data inform the final analysis. This approach could reduce power if there is a different treatment effect within the pooled treatment arms; however, the impact of this effect is expected to vary with the timing of the assessments within Phase II. The relationship between the design parameters and the operating characteristics will be assessed in Phase III. 

\subsubsection{Domain A Continued with Both Arms Retained}
If no arm is dropped, then the final analysis will also proceed using a gatekeeping approach. The primary analysis model including fluid arm and fludrocortisone arm will be constructed as below. Unlike the previous scenario, if neither of the newer fluid arms are dropping, then we will construct the following model where $I(X_{A_1 i}=1)$ and $I(X_{A_2 i}=1)$ represent indicator variables for if subject i was randomized to either of the newer fluids and $I(X_{B_1 i}=1)$ represents an indicator variable for if subject i was randomized to fludrocortisone.

$$logit(p_i )=\beta_0+\beta_1 I(X_{A_1 i}=1)+\beta_2 I(X_{A_2 i}=1)$$
$$+\beta_3 I(X_{B_1 i}=1)$$
To preserve $\alpha$, a gatekeeping procedure will be used to assess for any significant treatment effect among the fluid and fludrocortisone arms: 

$$H_{01}:\beta_1=\beta_2=\beta_3= 0 \hspace{2mm} vs.\hspace{2mm}  H_{a1}:\beta_1\neq 0|\beta_2\neq 0| \beta_3\neq 0$$

If the null hypothesis for the test above is rejected, then all pairwise combinations will be tested as below. 

$$H_{02}:\beta_1=\beta_2= 0 \hspace{2mm} vs.\hspace{2mm}  H_{a2}:\beta_1 \neq 0|\beta_2 \neq 0$$
$$H_{03}:\beta_1=\beta_3= 0 \hspace{2mm} vs.\hspace{2mm}  H_{a3}:\beta_1 \neq 0|\beta_3 \neq 0$$
$$H_{04}:\beta_2=\beta_3= 0 \hspace{2mm} vs.\hspace{2mm}  H_{a4}:\beta_2 \neq 0|\beta_3 \neq 0$$

Depending on which of the null hypotheses for the tests above are rejected, then individual tests comparing each arm to control will be completed. 

$$H_{05}:\beta_1=0 \hspace{2mm} vs.\hspace{2mm}  H_{a5}:\beta_1 \neq 0$$
$$H_{06}:\beta_2=0 \hspace{2mm} vs.\hspace{2mm}  H_{a6}:\beta_2 \neq 0$$
$$H_{07}:\beta_3=0 \hspace{2mm} vs.\hspace{2mm}  H_{a7}:\beta_3 \neq 0$$

For example, the null hypothesis $H_{05}$ will only be assessed if $H_{01}$, $H_{02}$, and $H_{03}$ are all rejected in their respective tests.

\subsubsection{Domain A Terminated at Feasibility Assessment}
In this scenario, after the feasibility assessment, randomization will cease within domain A. As such, for the final analysis will consider only the treatment assignments for Domain B.
For the final analysis in this scenario, subjects will be grouped as either having been randomized to fludrocortisone or no fludrocortisone. Therefore, let $I(X_{B_1 i}=1)$ represent an indicator variable for if subject $i$ was randomized to fludrocortisone, we can then construct the primary analysis model as below. 

$$logit(p_i)=\beta_0+\beta_1 I(X_{B_1 i}=1)$$

Under this scenario, only a single parameter $\beta_1$ is being assessed; therefore, no methods for the preservation of $\alpha$ are required.

\section{Simulation}
To evaluate the operating characteristics of the proposed design, specifically the impact of the timing of Phase II analysis on the overall operating characteristics of the trial, thorough simulation studies were conducted.  Parameters varied in the simulations include: timing of the feasibility assessment, timing of the arm-dropping assessment, effect of treatments in domain A on each biomarker of interest, effect of treatments in domain A on the Phase III outcome, and effect of treatments in domain B on the Phase III outcome.  

For timing of the feasibility and arm dropping analyses, the timing of each analysis is varied such that each can occur once outcomes from 90 subjects are available up to outcomes from 300 subjects, incrementing by 30. This allows for an assessment of both the effect of sample size for each analysis on the final operating characteristics and the effect that the order of the analyses has on the final operating characteristics. For each of the biomarkers associated with analyses in the Phase II portion of the trial, the impact of a treatment in domain A providing a mean change from baseline of either 0 or 10 units was assessed. For the Phase III portion of the trial, treatments in either domain can provide a 10\% change in the primary outcome ($Y_{21}$) relative to the control group. 
Simulations were conducted using R. For each simulation condition, 1,000 simulated datasets were generated. 

\section{Discussion}
Full simulation results can be seen in Appendix A and Appendix B. Figure 1  presents the probability of retaining the correct arm during the arm dropping analysis, the probability of proceeding beyond the feasibility analysis, and the overall probability of detecting the treatment effect across the possible sample size requirements for the feasibility and arm-dropping analysis for a representative set of simulation conditions (Scenarios A, B and C). Of note, the ordering of the feasibility and arm-dropping analyses is determined by their respective timings/required sample sizes. 

The graphs associated with row A represent the case where $Y_{11}:(CTL=0,A_1=-10,A_2=-10)$, $Y_{12}:(CTL=0,A_1=10,A_2=-10)$, and $Y_{21}: (A_1  =0.1,A_2 = 0.1,B_1=0)$. The opposing signs for the effect of treatment on $Y_{11}$ and $Y_{12}$ is clinically relevant for our motivating trial, as we expect one biomarker to increase the second to decrease with treatment; however, these opposing signs are inconsequential for the trial as both outcomes are continuous. In this case, the correct arm to retain during the arm-dropping analysis is $A_2$. This is a reflection of the the arm-dropping algorithm where the arm(s) with the highest average outcome $Y_{11}$ value and the lowest outcome $Y_{12}$ value being retained if there is a significant difference. If no difference is found, the the trial will proceed with $A_2$. Given that both $A_1$ and $A_2$ demonstrate the same mean shift from control for $Y_{11}$ and $Y_{12}$, then there is no expected difference between $A_1$ and $A_2$. 

The graphs associated with row B represent the case where $Y_{11}:(CTL=0,A_1=0,A_2=-10)$, $Y_{12}:(CTL=0,A_1=0,A_2=10)$, and $Y_{21}: (A_1 =0,A_2 = 0.1,B_1=0)$. Unlike Scenario A, Scenario B represents the case where $A_2$ demonstrates a difference from control when  $A_1$ is equivalent to control. In this case, the correct arm to retain during the arm-dropping analysis is $A_2$.

The graphs associated with row C represent the case where $Y_{11}:(CTL=0,A_1=-10,A_2=0)$, $Y_{12}:(CTL=0,A_1=10,A_2=0)$, and $Y_{21}: (A_1 =0.1,A_2 = 0.0,B_1=0)$. Similar to Scenario B, Scenario C represents the case where $A_1$ demonstrates a difference from control when $A_2$ is equivalent to control. As such, the correct arm to retain during the arm-dropping analysis is $A_1$.

\begin{figure*}[h!]
	\centering
	\includegraphics[width=1\linewidth]{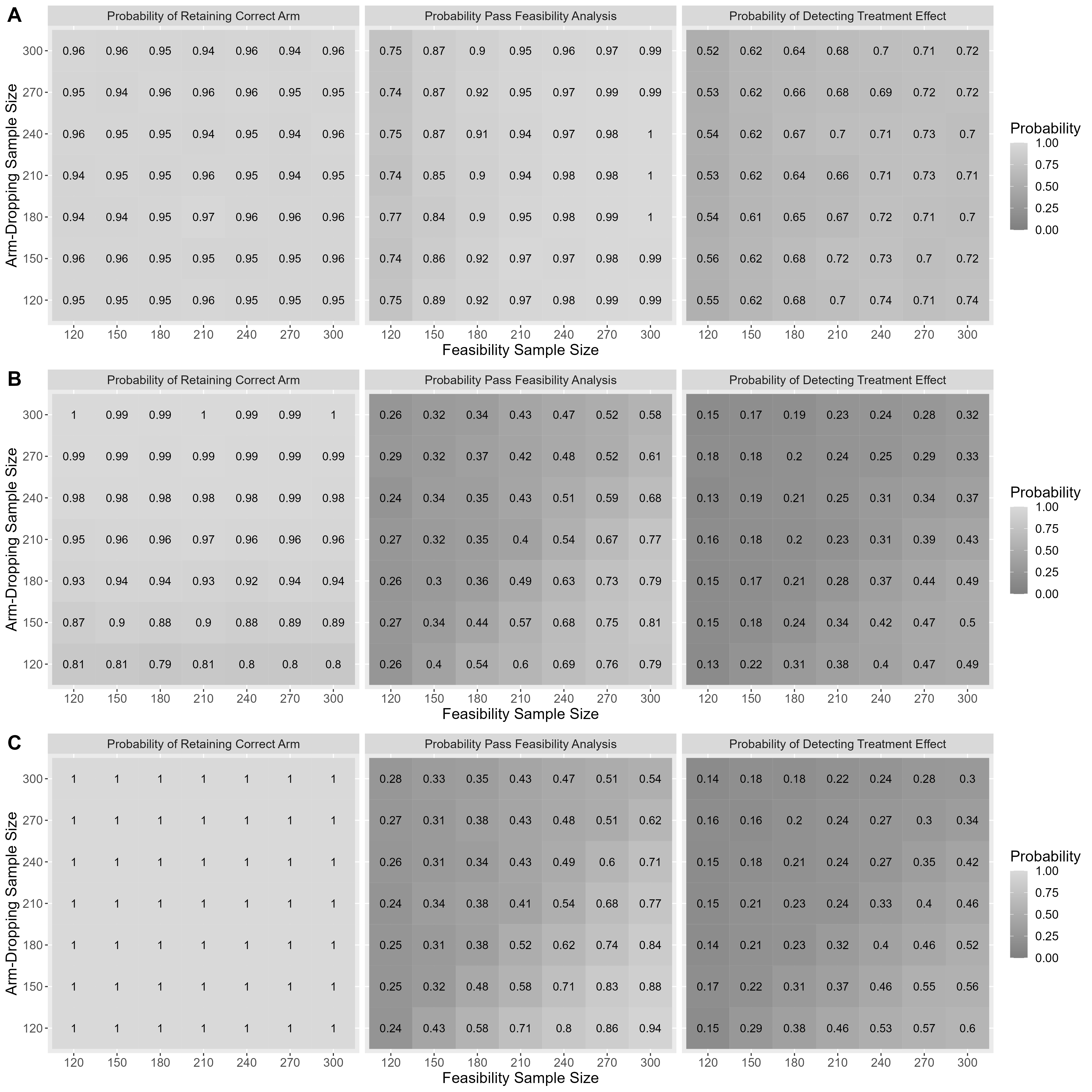}
	\caption{
		Figure 1: Simulation results for three scenarios (A,B,C). 
		The graphs associated with row A represent the case where $Y_{11}:(A_0=0,A_1=-10,A_2=-10)$, $Y_{12}:(A_0=0,A_1=10,A_2=-10)$, and $Y_{21}: (A_1  =0.1,A_2 = 0.1,B_1=0.0)$. 
		The graphs associated with row B represent the case where $Y_{11}:(A_0=0,A_1=0,A_2=-10)$, $Y_{12}:(A_0=0,A_1=0,A_2=10)$, and $Y_{21}: (A_1 =0,A_2 = 0.1,B_1=0.0)$. 
		The graphs associated with row C represent the case where $Y_{11}:(A_0=0,A_1=-10,A_2=0)$, $Y_{12}:(A_0=0,A_1=10,A_2=0)$, and $Y_{21}: (A_1 =0.1,A_2 = 0.0,B_1=0.0)$. }
	\label{fig:figure1}
\end{figure*}

Figure 1 shows that, for the arm dropping analysis, the probability of retaining the correct appears to be invariant to the sample size when the analysis is conducted for Scenarios A and C; however, in Scenario B the probability of retaining the correct arm increases with larger sample sizes. Further, it appears that the probability of retaining the correct arm is approximately constant for all feasibility analysis sample sizes.  

For the feasibility analysis, in general, the larger the sample size at the time of the analysis, the greater the probability of proceeding.  However, unlike the arm-dropping analysis, the timing of the arm-dropping analysis does appear to impact the results of the feasibility analysis. Specifically, when the sample size requirement is larger for the feasibility analysis, increasing the sample size required for the arm-dropping analysis appears to decrease the probability of proceeding to the Phase III portion of the trial. 

Lastly, we see that the power to detect the treatment effect in Phase III  is heavily dependent upon the timing of the feasibility and arm-dropping analyses. In general, overall study power appears to be maximized when the sample size for the feasibility analysis is large and when the sample size for the arm-dropping analysis is small. These results, which are similar to other conditions tested in simulation presented in Appendix A and Appendix B, indicate that the Phase II analyses should be ordered such that the arm-dropping analysis occur before the feasibility analysis.

\begin{figure*}[h]
	\centering
	\includegraphics[width=1\linewidth]{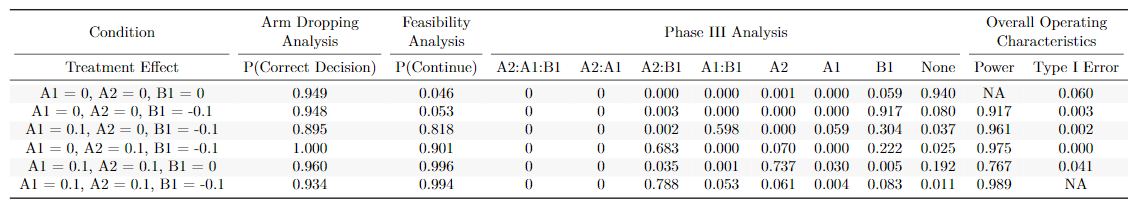}
	\caption{Table 1: Simulation results for the scenario listed in the column “Condition”. While "Condition" represents the effect of treatment on the primary Phase III outcome, for any treatment with a non-zero effect on the primary Phase III outcome, there is also a non-zero treatment effect on both Phase II outcomes. The columns grouped under the "Phase III Analysis" label represent the probability of declaring success for that collection of treatments. For example, in the last row under the column "A2:B1" the value is 0.788, which is interpreted as there being a 78.8\% chance of declaring success for A2 and B1 in this simulation scenario. The column "Type I Error" represents the family-wise Type I Error rate. }
	\label{fig:figure1}
\end{figure*}

Table 1 presents the characteristics across Phase III treatment effect conditions if the arm-dropping analysis were to occur once 150 outcomes are obtained and the feasibility analysis were to occur once 300 outcomes are obtained for scenarios where the arms which demonstrate an effect on the Phase III outcomes also demonstrate a treatment effect on both of the Phase II outcomes. Table 1 displays the likelihood of retaining the correct treatment arm (or both), the probability of continuing beyond the feasibility assessment, the probability of declaring each possible combination of treatment arms across domain as successful in Phase III, the overall study power, and the overall study Type I error. This table demonstrates that type I error (row $(A_1=0,A_2 = 0,B_1=0)$) is approximately 5\%. Further, we see that when at least one treatment arm within each domain has a non-zero effect, that power is expected to exceed 90\%. However, as expected, when there is a differential effect between the Phase II and Phase III outcomes or when only a single treatment arm is efficacious for both phases, then the overall study power decreases (as observed in Appendix A and Appendix B).

\section{Conclusions}
The class of FAST trial designs offers a framework which can provide increased efficiencies relative to standard and novel trial designs, such as a MAMS or factorial trial. Motivated by a trial currently under design, simulation studies demonstrated the numerous aspects of a FAST trial which need to be considered during the planning phase, such as the timing/sample size requirements for each assessment.

	\section*{Declarations}	
	
	\subsection*{Availability of data and materials}
	The results presented are derived from simulated data, as such they are not available. 
	
	\subsection*{Competing interests}
	The authors declare that they have no competing interests.
	
	\subsection*{Funding}
This work was supported by the National Institute of Neurological Disorders and Stroke of the National Institutes of Health under Award Number 5U01NS087748-09. The content is solely the responsibility of the authors and does not necessarily represent the official views of the National Institutes of Health.

	\subsection*{Author's contributions}
	JB, AM, and JE drafted the manuscript. All authors read and approved the final version of the manuscript.
	
	\subsection*{Acknowledgements}
	Not applicable.


\begin{thebibliography}{99}

\bibitem[\protect\citeauthoryear{Ghosh, Liu, and Mehta}{2020}]{Ghosh2020} 
Ghosh, P., Liu, L., Mehta, C.(2020). Adaptive multiarm multistage clinical trials{\it Statistics in Medicine}. { \bf 39,} 1084--1102.

\bibitem[\protect\citeauthoryear{Lin and Bunn}{2017}]{Lin2017} 
Lin, J., Bunn, V. (2019). Comparison of multi-arm multi-stage design and adaptive randomization in platform clinical trials{\it Contemporary Clinical Trials}. { \bf 54,} 48--59K.

\bibitem[\protect\citeauthoryear{Wason, Magirr, Law and Jaki}{2012}]{Wason2012} 
Wason, J., Magirr, D., Law, M., Jaki, T.(2012). Some recommendations for multi-arm multi-stage trials {\it Statistical Methods in Medical Research}. { \bf 25,} 716-727.


\bibitem[\protect\citeauthoryear{White et~al.}{2022}]{White2022} 
White, I., Petroni, G., Choodari-Oskooei, B., Sydes, M., Kahan, B., et al. (2022). Combining factorial and multi-arm multi-stage platform designs to evaluate multiple interventions efficiently. {\it Clinical Trials}. { \bf 19,} 432-441.

\bibitem[\protect\citeauthoryear{Jaki and Vasileiou}{2016}]{Jaki2016} 
Jaki, T., Vasileiou, T. (2016). Factorial versus multi-arm multi-stage designs for clinical trials with multiple treatments {\it Statistics in Medicine}. { \bf 36,} 563-580.

\bibitem[\protect\citeauthoryear{Sydes et~al.}{2009}]{Sydes2009} 
Sydes, M., Parmar, M., James, N., Clarke, N., Dearnaley, D., et al. (2009). Issues in applying multi-arm multi-stage methodology to a clinical trial in prostate cancer: the MRC STAMPEDE trial  {\it Clinical Trials}. { \bf 19,} 432-441.

\bibitem[\protect\citeauthoryear{Wang et~al.}{2009}]{Wang2009} 
Wang, D., li, Y., Wang, X., Liu, X., Fo, B., et al. (2009). Overview of multiple testing methodology and recent development in
clinical trials {\it Contemporary Clinical Trials}. { \bf 45,} 13--20.
		
\end{thebibliography}
\end{document}